%% file: spectralCheby_joint.tex
\pdfoutput=1
\documentclass[aps,prl,twocolumn,superscriptaddress,floatfix,citeautoscript,preprintnumbers]{revtex4-1}
\usepackage{graphicx,color}
\usepackage{amssymb}
\usepackage{amsmath}
\usepackage{epstopdf}
\usepackage[normalem]{ulem}
\usepackage{comment}
\usepackage{breakcites}
\usepackage[breaklinks=true]{hyperref}
\usepackage[caption=false]{subfig}
\usepackage{enumitem,floatrow}
\usepackage{soul}
\usepackage{bbold,dsfont}

\def\Id{\mathbb{1}}

\def\Z2{Z_{\mathrm{st2}}}

\newcommand{\beq}{\begin{equation}}
\newcommand{\eeq}{\end{equation}}
\newcommand{\bal}{\begin{align}}
\newcommand{\eal}{\end{align}}

\newcommand{\nn}{{\nonumber}}

\newcommand{\ket}[1]{\mbox{$ | #1 \rangle $}}
\newcommand{\bra}[1]{\mbox{$ \langle #1 | $}}
\newcommand{\ev}[1]{\langle #1 \rangle}
\newcommand{\opO}{\mathcal{ O }}
\newcommand{\geoM}{\mbox{$g_M( E ; \mathcal{ O } ) $ }}
\newcommand{\Xp}{\ket{X+}}
\newcommand{\Yp}{\ket{Y+}}
\newcommand{\Zp}{\ket{Z+}}
\newcommand{\Zz}{\ket{\mathcal{Z}_2}}
\newcommand{\Zzz}{\ket{\mathcal{Z}_3}}

\newcommand{\bed}{\[}
\newcommand{\eed}{\]}
\newcommand{\beqa}{\begin{eqnarray}}
\newcommand{\eeqa}{\end{eqnarray}}
\newcommand{\braket}[2]{\langle #1 | #2 \rangle}

\newcommand{\tr}{\mathop{\mathrm{tr}}}
 
\newcommand{\ie}{i.e.\ } % PR guide does not italicize

\definecolor{MyDarkBlue}{rgb}{0,0.08,0.45} 
\definecolor{MyLightMagenta}{cmyk}{0.1,0.8,0,0.1} 
\definecolor{MLM}{cmyk}{0.1,0.8,0,0.1} 
\definecolor{MyDarkGreen}{rgb}{0,0.45,0.08} 
\definecolor{MDG}{rgb}{0,0.55,0.05}

\usepackage[textsize=footnotesize]{todonotes}

\definecolor{atomictangerine}{rgb}{1.0, 0.6, 0.4}
\definecolor{bluegray}{rgb}{0.4, 0.6, 0.8}
\definecolor{brightube}{rgb}{0.82, 0.62, 0.91}
\definecolor{brilliantlavender}{rgb}{0.96, 0.73, 1.0}

\begin{document}

%\title{Probing thermalization with matrix product operators}
%\title{ Spectral analysis of matrix product operators } %% Favorite nr 2 ****
%\title{ Spectral analysis of quantum spin chains with matrix product operators }
%\title{ Spectral analysis of strongly correlated systems with matrix product operators }
%\title{Spectral properties of quantum many-body operators with tensor networks}
\title{Probing thermalization through spectral analysis with matrix product operators}

\begin{abstract}
	\input{abstract}
\end{abstract}

\author{Yilun Yang}
\affiliation{Max-Planck-Institut f\"ur Quantenoptik, Hans-Kopfermann-Str.\ 1, D-85748 Garching, Germany}
%\affiliation{Munich Center for Quantum Science and Technology (MCQST), Schellingstr. 4, D-80799 M\"unchen}
\author{Sofyan Iblisdir}
\affiliation{Departament de F\'{\i}sica Qu\`antica i Astronomia \& Institut de Ci\`encies del Cosmos, Universitat de Barcelona, Barcelona, Spain}
\affiliation{Departamento de An\'alisis y Matem\'atica Aplicada, Universidad Complutense de Madrid \& Instituto de Ciencias Matem\'aticas, Madrid, Spain}
\author{J. Ignacio Cirac}
\affiliation{Max-Planck-Institut f\"ur Quantenoptik, Hans-Kopfermann-Str.\ 1, D-85748 Garching, Germany}
\affiliation{Munich Center for Quantum Science and Technology (MCQST), Schellingstr. 4, D-80799 M\"unchen}
\author{Mari Carmen Ba\~nuls}
\affiliation{Max-Planck-Institut f\"ur Quantenoptik, Hans-Kopfermann-Str.\ 1, D-85748 Garching, Germany}
\affiliation{Munich Center for Quantum Science and Technology (MCQST), Schellingstr. 4, D-80799 M\"unchen}
% Activate to display a given date or no date
%\date{\today}							
\maketitle

\input{intro}
\input{geo} % g(E;O) as fundamental object; its applications
\input{method} % method and error sources
\input{models} % our numerical results
\input{thermalization}

\input{discussion}

\acknowledgments

This work was partly supported by the Deutsche Forschungsgemeinschaft (DFG, German Research Foundation) under Germany's Excellence Strategy -- EXC-2111 -- 390814868, and by the European Union through the ERC grants QUENOCOBA, ERC-2016-ADG (Grant no. 742102) \& GAPS (Grant no. 648913).  % and EU-QUANTERA project QTFLAG (BMBF grant No. 13N14780).
%%M.C.B. acknowledges the hospitality of KITP, where part of this work was developed, and support from the National Science Foundation under Grant No. NSF PHY-1748958.

\newpage
\appendix
\begin{center}
	\textbf{\large Supplemental Material}
\end{center}
\input{appendix}
\input{appendix-degenerate}

\bibliographystyle{apsrev4-1}
\bibliography{ChebyDOS}

\onecolumngrid
\end{document}

%% file: abstract.tex
% A first attempt
We combine matrix product operator techniques with Chebyshev polynomial expansions and present a method that is able to explore spectral properties of quantum many-body Hamiltonians. In particular, we show how this method can be used to probe thermalization of large spin chains without explicitly simulating their time evolution, as well as to compute full and local densities of states.
The performance is illustrated with the examples of the Ising and PXP spin chains. For the non-integrable Ising chain, our findings corroborate the presence of thermalization for several initial states, well beyond what direct time-dependent simulations have been able to achieve so far.

%% file: intro.tex
The study of one-dimensional quantum many-body systems has motivated the emergence of a number of techniques, based on tensor network states (TNS). More concretely, they use matrix product states (MPS) and matrix product density operators (MPDO)~\cite{Verstraete2008,Schollwoeck2011,Huckle2013tns,Orus2014a,Silvi2019tns} to approximate the ground states, low-lying excitations, thermal states, as well as time evolution. These methods have enabled the in-depth study of a multitude of models and the analysis of relevant physical phenomena. % \changeMC{[a Ref to rule them all]}. 

The success of such techniques is rooted in the ability of MPS and related ansatzes to accurately describe states that fulfill an area law of entanglement~\cite{Verstraete2006,Hastings2007}, satisfied (or only slightly violated) by many of the problems mentioned above~\cite{Eisert2010}. There are, however, important open questions that such techniques cannot easily solve. In particular, excited states at finite energy density are difficult to approximate, except in very particular cases~\cite{Khemani2016xdmrg,Kennes2015exc}, as they generically display volume law entanglement and, additionally, are embedded in highly dense spectral regions, which severely hinders the convergence of the algorithms. Out-of-equilibrium dynamics is also problematic: under time evolution a volume law often emerges, that makes an MPS approximation inadequate, except for short times. As a consequence, it is virtually impossible for standard MPS techniques to address the fundamental questions of equilibration and thermalization of relatively large closed quantum systems.

A  few alternative tensor network algorithms have tried to overcome these problems by avoiding the explicit representation of the states~\cite{Hartmann2009dmrg,Banuls2009fold,Enss2012lc,Kim2015slow}. Although they extend the applicability of the toolbox and allow access to additional dynamical quantities in some scenarios, the fundamental goal of accessing the long time behavior in a general case, and thus deciding the appearance of equilibration or thermalization, has not been achieved.

Here we introduce a new powerful tool to fill in these gaps. Our method is based on the use of MPO to approximate a family of generalized densities of states, and provides a means to directly address thermalization. More concretely, we combine TNS and  the kernel polynomial method (KPM)~\cite{Weisse2006kpm} in a general scheme that provides access not only to the full density of states (DOS) of a given many-body Hamiltonian~\footnote{TNS methods have been used to estimate DOS, but require to approximate the time evolution with MPS~\cite{Schrodi2017dos}.}, but also to energy functions that are intimately related to the out-of-equilibrium dynamics, including the local density of states (LDOS). With these functions it is possible to probe the eigenstate thermalization hypothesis (ETH)~\cite{deutsch91eth,srednicki94eth} across the spectrum, and to verify the thermalization of initial states \emph{without} explicitly simulating the time evolution. 

%Some earlier works have combined MPS with Chebyshev polynomials} ~\cite{Holzner2011cheMPS,Wolf2015cheby,Halimeh2015cheby} \st{, with a focus on the calculation of spectral functions and time-dependent quantities.  This approach is mostly related to our calculation in the particular case of the LDOS, but our method proposes a more general scheme and has much farther-reaching consequences. More recently,} \cite{Schrodi2017dos} \st{used TNS methods to estimate the DOS of a quantum many-body system, but the scheme was limited by the need to simulate the exponential evolution operator for long times.}

%% file: geo.tex
% 1) g(E,O) truncated to M as main object of interest
% 2) Particularizing O we get 
% 2.1. DoS -> thermodynamic properties (also other E-dependent ensembles!)
% 2.2. Microcanonical average op. (ratio)
% 2.3. LDOS => F(t) survival prob
% From GEO and LDOS, the diagonal ensemble

\paragraph{\textbf{Generalized DOS.---}}
Let us consider a quantum many-body Hamiltonian with spectral decomposition $H=\sum_{k} E_k \ket{k}\bra{k}$. We are interested in energy functions of the form 
\beq
g( E ; \opO ) = \sum_k \; \delta( E - E_k ) \; \bra{ k } \opO \ket{ k },
\label{eq:wdos}
\eeq
where $\opO$ is any operator and $\delta(x)$ is the Dirac delta function. We will aim an approximation to
\beq
g_M ( E ; \mathcal{ O } ) \equiv \tr \left[ \opO \delta_M(E-H)  \right],
\label{eq:wdosM}
\eeq
%\beq
%g_M( E ; \mathcal{ O } ) \equiv \sum_k \; \delta_M( E - E_k ) \; \bra{ k } \opO \ket{ k }=\tr \left [ \delta_M(E-H) \opO \right ] / d_{\mathcal{H}},
%\label{eq:wdosM}
%\eeq
where $\delta_M$ is a smooth function such that $\lim_{M \to \infty} \delta_M = \delta$. As we will show below, \geoM can be computed from traces $\tr \left[ \opO T_n(H)\right]$, where $T_n(H)$ are the Chebyshev polynomials of $H$~\footnote{Earlier works have combined MPS and Chebyshev polynomials to approximate the dynamics and to compute spectral functions ~\cite{Holzner2011cheMPS,Wolf2015cheby,Halimeh2015cheby}.}.

%Before describing the method we use to approximate $g(E;\opO)$, it is interesting to discuss its implications for physically relevant quantities.
Being able to estimate $g_M(E;\opO)$ allows us to access a number of physical quantities that we can use to probe the dynamics of $H$:
\begin{enumerate}[label=(\roman*)]
\item \label{enum:dos}
	$g_M(E;\Id) / d_{\mathcal{H}}$ is a broadened DOS, where $d_\mathcal{H}$ is the dimension of the Hilbert space. It thus enables the computation of thermodynamic quantities. For instance, the partition function in the canonical ensemble can be computed as $Z_M({\beta})=\int dE e^{-\beta E} g_M(E;\Id)$.
\item \label{enum:Omc}
	Since Eq.~\eqref{eq:wdos} represents the (unnormalized) average expectation value of $\opO$ over all states with the same energy, the expectation value of $\opO$ in the microcanonical ensemble $O(E)$ is given by the ratio~\footnote{Provided the DOS does not vanish; otherwise the microcanonical ensemble does not have any component at that $E$.}
	%\note[color=atomictangerine]{And if it vanishes, there is no microcanonical ensemble component at that $E$.}
	\beq
	O(E)=\frac{g(E;\opO)}{g(E;\Id)}\approx \frac{g_M(E;\opO)}{g_M(E;\Id)}\equiv O_M(E).
	\label{eq:Omc}
	\eeq
\item \label{enum:ldos}
	If the operator is taken to be a projector onto a pure state $\opO=\ket{\Psi}\bra{\Psi}$, the computed function, which we denote  $g_M(E;\Psi)=g_M(E;\ket{\Psi}\bra{\Psi})$, is the corresponding LDOS. 
	%and encodes information about the evolution of the state under $H$ at arbitrarily long times. 
\end{enumerate}

\paragraph{\textbf{Dynamical probes.---}}
%Using these quantities we can probe the dynamics in several ways.
Firstly, using~\eqref{eq:Omc} we can probe some of the predictions of ETH~\cite{deutsch91eth,srednicki94eth,Srednicki99,Rigol2008therm,DAlessio2016eth},
%the microcanonical estimate~\ref{enum:Omc} can be used to probe ETH. The latter
which postulates that, regarding physical observables~\footnote{Physical observables include in particular (but are not restricted to) local ones. See~\cite{DAlessio2016eth} for a discussion.}, energy eigenstates \emph{look thermal}, \ie they have expectation values close to those of an equilibrium ensemble with a temperature set to get the same mean energy
\footnote{More specifically, ETH postulates~\cite{Srednicki99} that in the energy basis physical observables have matrix elements of the form: $O_{kq} = O(\bar{E})\delta_{kq}+e^{-{S(\bar{E})/2}} f(E,\Delta E) R_{kq}$, where $\bar{E}$ and $\Delta E$ are the mean and difference of eigenvalue energies, and $O$ and $f$ are smooth functions of their arguments. Furthermore, $O(\bar{E})$ is the expectation value of the observable in the microcanonical  ensemble at energy $\bar{E}$, and $R_{kq}$ is a random variable with zero mean and unit variance. $S(E)$ is the thermodynamic entropy, defined in as $e^{S(E)}=E \sum_k \delta_{\epsilon}(E-E_k)$, where $\delta_{\epsilon}$ is a broadened $\delta$ function, such that $S(E)$ is monotonic.}.
If ETH holds we thus expect the estimate $O_M(E)$ to be a smooth function of energy, and to be equal to the thermal value at the same mean energy,
\beq
O_M(E)\stackrel{\mathrm{ETH}}{\approx}
\tr\left[  \rho_{\beta(E)} \opO \right ],
\label{eq:ethProbe}
\eeq
%\note[color=atomictangerine]{Still mismatch micro- vs canonical.}
where $\beta(E)$ is the corresponding temperature. Probing this relation constitutes a weak test of ETH. Secondly, we can use the estimates \ref{enum:Omc} and \ref{enum:ldos} to approximate the long-time averaged expectation value $\bar{O}=\lim_{T\to \infty}\frac{1}{T}\int_0^T dt \bra{\Psi(t)} \hat{ O } \ket{\Psi(t)}$ which, if the spectrum is not degenerate, 
is given by the expectation value in the diagonal ensemble, $\bar{O}=O_{\mathrm{Diag}}(\Psi)\equiv \sum_k \bra{k} \opO \ket{k} |\bra{k}\Psi\rangle|^2$. Under the non-degeneracy condition, $\bra{k} \opO \ket{k}=O(E_k)$, and we can write
\begin{align}
O_{\mathrm{Diag}}(\Psi)&=\sum_k \int dE \delta(E-E_k)  O(E)  |\bra{k} \Psi\rangle|^2 \nn \\
 &= \int dE O(E) g(E;\Psi) \approx \int dE O_M(E) g_M(E;\Psi),
 \label{eq:diagEV}
\end{align}

If the system thermalizes, the long-time value will be thermal, so we expect
\beq
\int dE O_M(E) g_M(E;\Psi) \approx
 \tr\left[  \rho_{\beta(E)} \opO \right ],
\label{eq:thermalization}
\eeq
for $\ev{E}=\bra{\Psi}H\ket{\Psi}$. Hence it is possible to probe the thermalization of individual initial states without the need to explicitly simulate time evolution. Instead, we can estimate, as we detail in the following, the expectation value of any local observable in the diagonal ensemble for initial states that can be written as an MPS, and compare this result to the expectation value in the Gibbs ensemble for which the mean energy is $\langle E \rangle$ (which for local Hamiltonians can be efficiently approximated using MPS tools). Notice that if the energy spectrum has degeneracies, it is still possible to estimate the long-time averaged $\bar{O}$ with our method, and perform this comparison, although with a higher computational cost~\cite{SuppMat}.

Finally, the LDOS encodes information about the evolution of a state under $H$ at arbitrarily long times. Indeed, its Fourier transform  (assuming $H$ is constant in time) is the survival probability, $F(t)\equiv \left|\bra{\Psi(0)} \Psi(t)\rangle\right|^2$ which is sensitive to all time regimes of the evolution~\cite{Santos2018}, 
\beq
F(t)=\left| \int dE e^{-iE t} g(E;\Psi)\right|^2 \approx \left| \int dE e^{-iE t} g_M(E;\Psi)\right|^2.
\label{eq:surv}
\eeq
The decay of the survival probability after a quench presents different regimes, and shows sensitivity towards ergodicity and thermalization~\cite{Santos2018,Tavora2016power,TorresHerrera2015mbl,Schiulaz2019thouless}. %Moreover it has been recently used to characterize the dynamics near the MBL transition~\cite{}, and to identify long timescales in interacting systems~\cite{}.

%% file: method.tex
\paragraph{ \textbf{Chebyshev expansions.---} }
%%%%%%%%%%%%%%%%%%%%%%%%%

The basis of our numerical strategy is the expansion of the Dirac delta function in terms of Chebyshev polynomials $T_n$, defined by the recurrence relation~\cite{Weisse2006kpm}
\begin{align}
	&T_0( x ) = 1, \; T_1( x ) = x,  \nn \\
	&T_{ n + 2 }( x ) = 2 x \; T_{ n + 1 }( x ) - T_n( x ), \; n > 0.
	\label{eq:chebyshev-recurrence}
\end{align}
Any piecewise continuous function $f(x)$ with $x\in[ -1, +1]$ admits such expansion~\cite{Pinkus2000,Boyd2000,Weisse2006kpm}, and can be approximated by a truncated sum:%~\footnote{The modified expansion in terms of $T_n(x)/\sqrt{1-x^2}$ is used here for convenience, instead of the linear combination of polynomials (see~\cite{Weisse2006kpm}).}
\beq\label{eq:chebyshev-approx}
f( x ) \approx \frac{1}{ \pi \sqrt{ 1 - x^2 } } \big[ \gamma_0 \mu_0 + 2 \sum_{ n = 1 }^{ M - 1 } \gamma_n \mu_n T_n( x ) \big].
\eeq
%{\color{blue} A short exposition of the properties of Chebyshev polynomials most relevant to this work can be found in supplemental material \cite{SuppMat}.}  
The moments $\mu_n = \int_{-1}^{1} f( x ) T_n( x ) dx$ are the coefficients of the full expansion,
while the $\gamma_n$ are introduced by the KPM to improve the quality of the truncated approximation,
and depend on the order of the truncation $M$, but not on $f$~\cite{SuppMat}.
 
Using the expansion for the delta function~\cite{Weisse2006kpm} for each term in \eqref{eq:wdos}, \geoM can be written in the form \eqref{eq:chebyshev-approx}, with moments
%\bed
%\mu_n( H; \mathcal{O})=\frac{1}{\nu d_{\mathcal{H}}}\sum_k\int_{-1}^{1} dx \delta(x-\varepsilon_k) \bra{k} \opO \ket{k} T_n(x)
%\eed
\beq
\mu_n( H; \mathcal{O}) \equiv  \frac{1}{\nu}\tr \left[ \opO T_n(\tilde{H}) \right],
\eeq
where $\tilde{H}=H/\nu+\Delta E$ is the rescaled and potentially shifted Hamiltonian, such that the spectrum $\varepsilon_k=E_k/\nu+\Delta E$ is strictly contained in $[-1,\,1]$~\cite{SuppMat}.

We can construct fixed bond dimension MPO approximations to the polynomials $T_n^{(D)}(\tilde{H})\approx T_n(\tilde{H})$ for any Hamiltonian $H$ that is itself expressed as an MPO. Starting from $T_0(\tilde{H})=\Id$ and $T_1(\tilde{H})=\tilde{H}$ (both exact MPO), we apply the recurrence relation between Chebyshev polynomials~(\ref{eq:chebyshev-recurrence}). This increases the bond dimension, so at each step we approximate the result with the maximum $D$ allowed using standard TNS techniques~\cite{Verstraete2008}, $T_{ n + 2 }^{(D)}( \tilde{H} ) \approx 2 \tilde{H} T^{(D)}_{n+1}(\tilde{H}) - T^{(D)}_{n}(\tilde{H})$. We can then compute the traces $\tr [ \opO T_{n}^{(D)} ]$ and thus approximate the function \geoM  for any operator $\opO$ which can also be expressed as an MPO.
%\footnote{It is also possible to promote $\geoM$ to an operator, $g_M(H;\opO)\equiv \sum_k g_M(E_k;\opO) \ket{E_k} \bra{E_k}$, 
%and approximate it by a MPO using the strategy described in the appendix for the step function, and directly approximating the sum of 
%polynomials with a MPO. Although this may have its interest 
%for some applications, we find that the truncation errors accumulate fast with the order of the expansion. }

The case of the LDOS allows for a more efficient implementation, since in that case the traces to be evaluated reduce to the single expectation value $\bra{\Psi} T_n^{(D)}(\tilde{H}) \ket{\Psi}$. Thus, instead of each full polynomial, it is enough to find an MPS approximation of the vectors resulting from applying them, $\ket{t_n}\equiv T_n(\tilde{H})\ket{\Psi}$, which satisfy the same recurrence relation. This reduction of Chebyshev expansions to states was used in \cite{Holzner2011cheMPS} to estimate spectral functions.

The sources of errors in our approach are analyzed in the supplemental material~\cite{SuppMat}. Note that because of the largely varying DOS across energy regions (in particular for local models as considered here, the DOS is Gaussian in the thermodynamic limit~\cite{Hartmann2005dos,Keating2015dos}), the precision of \geoM estimated with the above procedure worsens near the edges of the spectrum as discussed below. We can alleviate this problem by applying separate expansions to the Hamiltonian projected onto the different energy intervals, $H \to \theta(H-E_{\mathrm{cut}})  H$ (for high) or $\theta(E_{\mathrm{cut}}-H) H$ (for low energies)~\footnote{Using a similar strategy, it is also possible to project inside a closed interval, $\theta(H-E_{\mathrm{low}}) \theta(E_{\mathrm{high}}-H) H$.}. Since the step function $\theta(x)$ can also be approximated using the KPM, this construction can be realized within our numerical method~\cite{SuppMat}.

%% file: models.tex
\paragraph{\textbf{Models.---}}
We have applied the method to %Chebyshev approximation presented above %to compute the spectral properties for 
two quantum spin chains %with interesting dynamics, both 
with open boundary conditions (the scheme can be also used for periodic chains). The first is the Ising model,
\beq
H_{\mathrm{Ising}}=J\sum_{i=1}^{N-1} \sigma_z^{[i]} \sigma_z^{[i+1]}  + g \sum_i^N \sigma_x^{[i]}  + h \sum_i^N \sigma_z^{[i]} , 
\label{eq:Hising}
\eeq
in general non-integrable, except in the limits $g=0$ (classical) or $h=0$ (transverse field Ising model). This Hamiltonian has been profusely studied in the context of quantum quenches. Non-trivial dynamics has been observed and investigated in the non-integrable regime~\cite{Banuls2011therm,Hastings2015impro,Kormos2016nat,Lin2017quasip,James2019conf}, in particular, for the parameters that we consider, $(J, g, h)=(1,-1.05,0.5)$. For comparison, we analyze also the integrable point $(1,0.8,0)$.

Second, we consider the PXP model,
\beq
H_{\mathrm{PXP}}=\sum_{i=2}^{N-1} P_{i-1} \sigma_x^{[i]} P_{i+1} +  \sigma_x^{[1]} P_2 + P_{N-1}  \sigma_x^{[N]},
\label{eq:Hpxp}
\eeq
where $P_i=(1-\sigma_z^{[i]})/2$. This kinetically constrained model was recently realized in a Rydberg atom chain experiment~\cite{Bernien2017}, and the observation of persistent revivals for particular initial configurations has triggered intense theoretical investigation about quantum scars as a possible mechanism to prevent thermalization~\cite{Turner2018scars,Turner2018prb,Lin2019scars,Khemani2019pxp,Pichler2019scars}.

%% file: thermalization.tex
\begin{figure}
	\centering
	\begin{minipage}{.494\columnwidth}
		\includegraphics[width=\columnwidth]{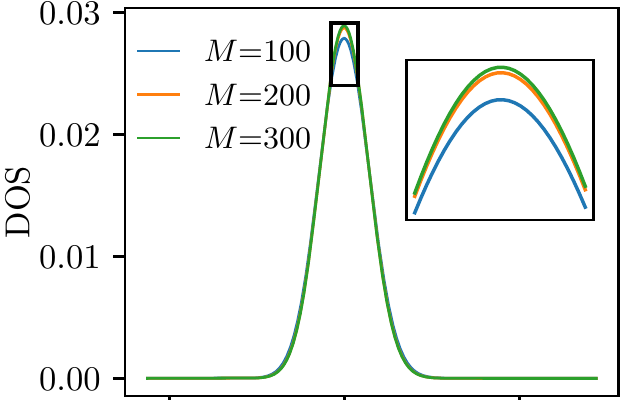}
	\end{minipage}
	\begin{minipage}{.494\columnwidth}
		\includegraphics[width=\columnwidth]{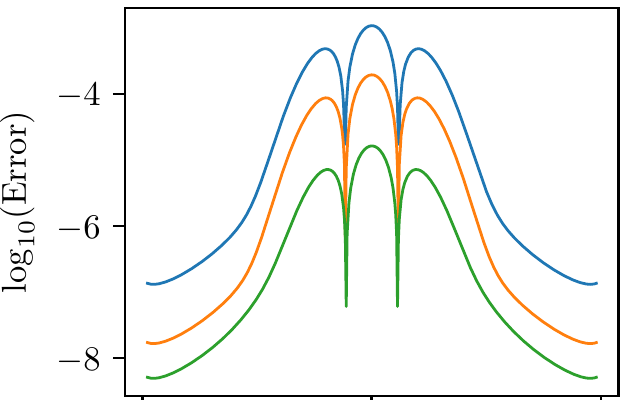}
	\end{minipage}\\
	\begin{minipage}{.494\columnwidth}
		\includegraphics[width=\columnwidth]{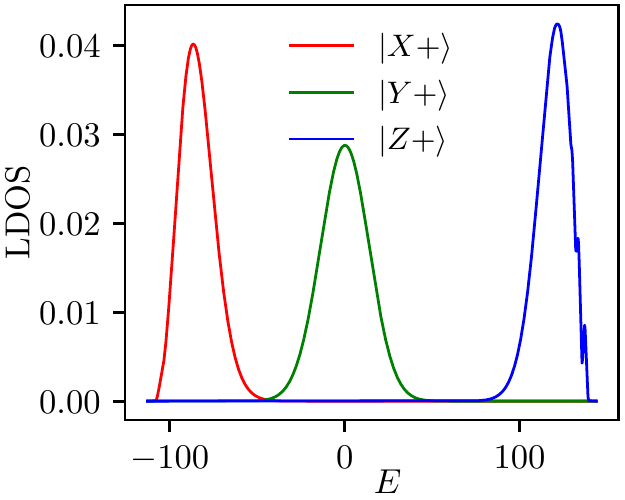}
	\end{minipage}
	\begin{minipage}{.494\columnwidth}
		\includegraphics[width=\columnwidth]{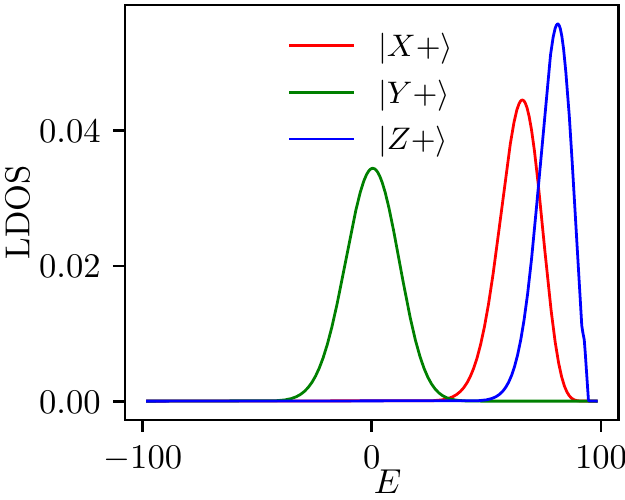}
	\end{minipage}
	\caption{
		Results of the Ising chains for $N=80$, using bond dimension $D=200$. The upper panels show, as a function of energy, the DOS for the non-integrable case (left) and the error of DOS in the integrable one (right) for several values of the truncation order $M$. The lower panels show the LDOS in the non-integrable (left) and integrable (right) case for totally polarized initial states \Xp, \Yp, \Zp.
	}
	\label{fig:ising_dos}
\end{figure}

\begin{figure}
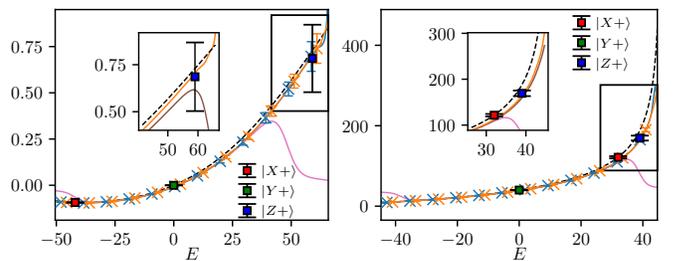

\centering
\begin{minipage}{.494\columnwidth}
\includegraphics[width=\columnwidth]{{{02_01_Ising_thermal_J1gm105h05_L40_SzMid}}}
\end{minipage}
\begin{minipage}{.494\columnwidth}
\includegraphics[width=\columnwidth]{{{02_02_Ising_thermal_J1g08h0_L40_Sz2}}}
\end{minipage}
\caption{
	Thermalization probes for the non-integrable (left) and integrable (right) Ising models, for a chain of $N=40$ sites. Dashed black line: thermal expectation value of a particular operator. Orange line: $O_M(E)$~\eqref{eq:Omc} with $\theta$ projections; $M=100$ (left), resp. $M=150$ (right); error bars indicate the difference with respect to truncation $M-50$ (brown line in the inset); $D=600$ (blue line for $D=200$, with negligible error from bond dimension effect). Pink line: $O_M(E)$ of same $M$ without $\theta$ projections, failing for high energy regions. The red, green and blue points show the diagonal expectation value~\eqref{eq:diagEV} for the different initial states. 
}
\label{fig:Ising_therm}
\end{figure}

\begin{figure}[h]
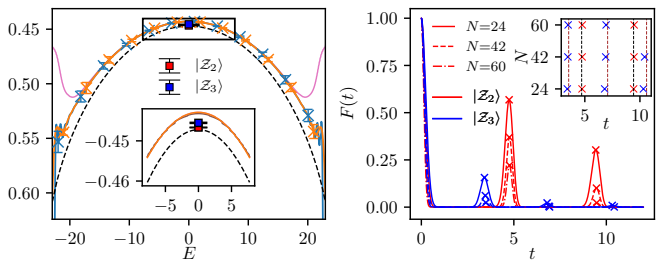

\centering
\begin{minipage}{.494\columnwidth}
\includegraphics[width=\columnwidth]{{{03_01_PXP_thermal_L40_SzMid}}}
\end{minipage}
\begin{minipage}{.494\columnwidth}
\includegraphics[width=\columnwidth]{{{03_02_PXP_SurvivalProb}}}
\end{minipage}
\caption{
	Thermalization probes in the PXP model. Left: relation~\eqref{eq:ethProbe} holds only in the center (colors as in fig.~\ref{fig:Ising_therm}); for initial states $\Zz$, $\Zzz$ (red and blue symbols), at $E=0$, \eqref{eq:diagEV} agrees with the thermal value. Right: survival probability of both states as a function of time for different sizes. Revival times are almost independent of system size, and agree with predictions in~\cite{Turner2018prb} (dashed lines in the inset).
}
\label{fig:PXP_th}
\end{figure}

\paragraph{ \textbf{ Thermalization probes.---} }
%%%%%%%%%%%%%%%%%%%%%%%%

To probe thermalization in the Ising model, we consider three initial states that we call \Xp, \Yp, \Zp, defined as translationally invariant products of totally polarized spins in the corresponding directions. Their LDOS for the integrable and non-integrable Ising models, and the DOS of both Hamiltonians are shown in Fig.~\ref{fig:ising_dos} for a chain of $N=80$ particles. The results for the DOS are very precise already for moderate bond dimensions $D$ and truncation parameter $M$.

In the non-integrable case we analyze the thermalization of $\opO=\sigma_z^{[N/2]}$. In the integrable one, this operator vanishes, and instead we consider $\opO=(\sum_i \sigma_z^{\left[ i \right]})^2$. 
%\note[color=atomictangerine]{Though physical, the $\opO$ for the integrable case is not local, so finite size effects, at least, would be more important.} 
The results are shown in Fig.~\ref{fig:Ising_therm} for a chain of $N=40$ sites. To check relation~\eqref{eq:ethProbe}, we plot $O_M(E)=\geoM/g_M(E;\Id)$ (in yellow) as a function of $E$, and at the same time the value in the corresponding Gibbs ensemble, i.e. such that $E=\tr (H e^{-\beta H}) /Z$ (dashed black line). In the non-integrable case we observe convergence within the error bars (estimated from the comparison between different truncation orders). In the integrable one, we observe a deviation in the region of largest energy. But in this case there may be eigenstates that do not fulfill ETH, and the relation (\ref{eq:ethProbe}) does not need to hold. Notice that if we had not used different ($\theta$-projected) Chebyshev expansions for different energy sectors, the results do not converge in the outer parts of the spectrum (pink lines in the figures).

We also probe thermalization for the initial states mentioned above by checking relation~\eqref{eq:thermalization}. For each of the \Xp, \Yp and \Zp states, the figures show the result of evaluating the RHS of~\eqref{eq:diagEV} for the observable $\opO$ analyzed in the corresponding model versus $E=\bra{\Psi}H\ket{\Psi}$. If the state thermalizes, and the approximation~\eqref{eq:diagEV} is good enough, we expect that the result agrees with the thermal expectation value (black curve) at the same mean energy.

In the non-integrable case (left panel of Fig.~\ref{fig:Ising_therm}), the error bars are compatible with thermalization for the three states. This is particularly interesting for the $\ket{X+}$ and $\ket{Z+}$ states, for which numerical simulations are not able to reach thermalization times~\cite{Banuls2011therm,Hastings2015impro}, but there are arguments for eventual thermalization~\cite{Lin2017quasip}. The significantly larger error bar for the $\ket{Z+}$ state is related to the closeness of this state to the edge of the spectrum, which makes it sensitive to the discrete character of the latter, as evidenced also in the corresponding LDOS (lower left panel in fig.~\ref{fig:ising_dos}). In the integrable case, the value of the most energetic of the states, $\ket{Z+}$, is not compatible with the assumption of thermal equilibrium, even with error bars. 
In this case, if the system equilibrates, we expect it to be to a generalized Gibbs ensemble ~\cite{Vidmar2016gge}. Nevertheless, our observation cannot be taken as a test of such effect, because the estimate~\eqref{eq:diagEV}, in a case with degeneracies in the spectrum, does not necessarily correspond to the expectation value in the long time limit~\cite{SuppMat}.

%A different perspective on the potential thermalization of the individual states is provided by the time-dependent survival probability~\eqref{eq:surv}. {\color{blue}We analyse it here for the PXP model, and in the appendix for the Ising models. Fig.~\ref{fig:PXP_th} shows this probability for a system of $40$ spins.}

A similar analysis for PXP model is shown in Fig.~\ref{fig:PXP_th} (left panel) for a system of $40$ spins. We compare the microcanonical estimate~\eqref{eq:Omc} for the operator $\opO=\sigma_z^{[N/2]}$ (yellow line and symbols) to the thermal value (black line), and observe that the agreement is best close to the center of the spectrum, but values increasingly differ (error bars considered) towards the edges. This observation is compatible with exact diagonalization results~\cite{Turner2018prb} (for much smaller systems) that predict the existence of ETH-violating eigenstates (scar states) in all regions of the spectrum. Closer to the edges of the spectrum, the ratio of scar states with respect to ETH ones becomes non-negligible, which explains the more evident breaking of ETH in these regions.

We next consider two initial states $\Zz \equiv\ket{\uparrow \downarrow \uparrow \downarrow \ldots}$ and $\Zzz \equiv\ket{\uparrow \downarrow \downarrow \uparrow \downarrow \downarrow \ldots}$, for which unexpectedly long-lived oscillations have been experimentally observed in Rydberg atoms~\cite{Bernien2017}. It has been recently postulated~\cite{Turner2018scars,Turner2018prb,Lin2019scars,Khemani2019pxp} that the slow dynamics of these states is due to their large overlap with a few scar states. These states lie nevertheless in the middle of the spectrum, $E=0$, where there is an exponentially large degeneracy, so that Eq.~(\ref{eq:diagEV}) (shown in the figure and compatible with the thermal value within error bars) is not necessarily a good estimate of the long-time limit.
 
The survival probability of these states, in contrast, does show the peculiarities of the real time dynamics of these two states. As shown in the right panel of Fig.~ \ref{fig:PXP_th}, the fidelities of both states show periodic revivals. 
We observe that for both $\Zz$ and $\Zzz$, the height of the peaks seems to decrease exponentially with the system size, which we choose to be multiples of 6~\cite{Turner2018prb}. The locations of the peaks are more robust (see inset), at times $t_{\mathcal{Z}_2}\approx3\pi n /2$ and $t_{\mathcal{Z}_3}\approx9\pi n/8$, for $n \in \mathbb{Z}$, in agreement with the prediction in~\cite{Turner2018prb}.

%\mycomment{MC}{After reading some of the refs, in particular\cite{Tavora2016power,Santos2018} I think we can say something more about the SP figures: the final drop is a consequence of finite $M$ (as we see from the difference). Makes sense: we need resolution of smaller energy differences to reach longer times.
%The initial drop is predicted to be Gaussian, due to the Gaussian LDOS  in all cases, with $\sigma^2$ the variance of the LDOS. Then, in chaotic systems, and for systems in the middle of the spectrum, a power-law behavior appears (envolvent of the oscillations), with $1/t^2$ for disordered spin chains, as we seem to see for $\ket{Y+}$ in the non-integrable case (could we super impose a $1/t^2$ or $1/t^3$ -- this is for fully random matrices-- line on Fig. \ref{fig:isingNI_surv}?). 
%The power law is due to the finiteness of the spectrum and it is slower for 
%states closer to the edges (we do not really see a decay in $\ket{X+}$ and $\ket{Z+}$).
%In finite systems, it saturates, but before that, the appearance of a \emph{correlation hole} (dip below saturation) indicates level-repulsion.
%We are not seeing either saturation or correlation hole, because $M$ truncation hits earlier. Actually, from the onset of the error, we
%can: 1) predict that the required $M$ grows exponentially with the time we want (surprise, surprise); 2) say that the results are reliable for intermediate times (almost $t~100$). Also, I would redo the figures for SP cutting the lines after the error starts (remove the final drop, which is not reliable).}

%% file: discussion.tex
\paragraph{{\textbf Discussion.---}}
We have presented a technique, based on Chebyshev expansions and MPS algorithms, to compute generalized densities of states, and have shown how it can be used to directly probe thermalization in one-dimensional quantum many-body models.

We consider a broad range of potential extensions and applications to be analyzed in the future. First of all, our calculations for spin models can be easily extended to disordered, quasi-periodic, long-range interacting, bosonic, fermionic and even two-dimensional systems. Beyond Hamiltonians, our scheme carries over to any sort of MPO, and could be useful to explore the spectral properties of Lindbladians, random MPO, or others. Finally, the scheme used to compute the survival probability can also be extended to monitor the evolution of (local) observables, even at finite temperature, and thus provide new tools to study the fundamental questions of out-of-equilibrium dynamics.

%% Outlook
%The present work contributes to 
%using 
%approximate polynomial expansions (APE) of matrix product operators 
%for the investigation of quantum many-body physics,
%a route initiated.... %% ** citar primeros trabajos, y August2018mpo
%We believe that the potential of these techniques is far from exhausted,
%and conclude this paper by mentioning a few applications to be 
%analyzed in the future.
%
%Density of state calculations can be straightforwardly extended to 
%disordered, quasiperiodic, long-range interacting, bosonic, fermionic and 
%even two-dimensional systems.
%Then, we would have access to all thermodynamic properties thereof.
%Of particular interest could be the application to lattice gauge theories~\cite{Langfeld2012dos},
%for which tensor network techniques have proven to be useful~\cite{Dalmonte2016lgt,Banuls2018lat}.
%Beyond Hamiltonians, 
%our scheme carries over to any sort of MPO, including Lindbladians, random MPO, or others.
%
%%Probing the validity of ETH requires testing not only the average expectation values, as we 
%%have shown, but also the behavior of off-diagonal matrix elements.
%%We can envision extensions of the spectral techniques to examine this question.
%
%Finally, the scheme used to compute the survival probability can also be extended
%to monitor the evolution of (local) observables, even at finite temperature, 
%and thus provide new tools to study 
%the fundamental question of out-of-equilibrium dynamics.

%\changeMC{Say something about potential improvements with energy truncation? Alternative to $\theta$. At least in appendix.}

%% file: appendix.tex
\section{Two types of Chebyshev expansions\cite{Weisse2006kpm}}
%%%%%%%%%%%%%%%%%%%%%%%%%%%%%%%%%%%%

In this letter, we are using the Chebyshev polynomials of the first kind, $T_n(x)$,
%{\color{blue} They are defined by the recurrence relation
%\begin{align}
%&T_0( x ) = 1, \; T_1( x ) = x,  \nn \\
%&T_{ n + 2 }( x ) = 2 x \; T_{ n + 1 }( x ) - T_n( x ), \; n > 0.
%\label{eq:chebyshev-recurrence}
%\end{align}}
and two different ways to expand a function in terms of $T_n$. Consider the two inner products of functions $f(x)$ and $g(x)$ on $[-1,1]$:
\begin{eqnarray}
	\begin{aligned}
		{\braket{f}{g}}_1 & = \int_{-1}^{1} \frac{f(x)g(x)}{\pi \sqrt{1-x^2}}dx,\\
		{\braket{f}{g}}_2 & = \int_{-1}^{1} \pi \sqrt{1-x^2}f(x)g(x)dx.\\
	\end{aligned}
\end{eqnarray}
The orthogonality relations of $T_n(x)$ follow as
\begin{eqnarray}
	{\braket{T_n}{T_m}}_1 = {{\braket{\phi_n}{\phi_m}}}_2 = \frac{1+\delta_{n,0}}{2}\delta_{n,m},
\end{eqnarray}
where $\phi_n(x)=\frac{T_n(x)}{\pi\sqrt{1-x^2}}$. Thus the Chebyshev expansion can be given by either
\bed
f(x)=\sum\limits_{n=0}^{\infty}\frac{{\braket{f}{T_n}}_1}{\braket{T_n}{T_n}_1}T_n(x)=\alpha_0 + 2\sum\limits_{n=1}^{\infty}\alpha_n T_n(x),
\eed
where $\alpha_n=\braket{f}{T_n}_1=\int_{-1}^{1}\frac{f(x)T_n(x)}{\pi\sqrt{1-x^2}}dx$, or
\begin{eqnarray}
	\begin{aligned}
		f(x) & =\sum\limits_{n=0}^{\infty}\frac{{\braket{f}{\phi_n}}_2}{\braket{\phi_n}{\phi_n}_2}\phi_n(x)\\
		& =\frac{1}{\pi\sqrt{1-x^2}}\left[\mu_0 + 2\sum\limits_{n=1}^{\infty}\mu_n T_n(x)\right],
	\end{aligned}
\end{eqnarray}
where $\mu_n=\braket{f}{\phi_n}_2=\int_{-1}^{1}f(x)T_n(x)dx$. As a consequence of the Stone-Weierstrass theorem \cite{Pinkus2000,Boyd2000}, any continuous function on $[-1,+1]$ admits a converging expansion in terms of Chebyshev polynomials.

In practice, if we cut off at a finite number $M$ of Chebyshev terms, there would show Gibbs oscillations near the regions where the function is not continuously differentiable. 
The kernel polynomial method suppresses these oscillations by introducing kernels, \emph{i.e.}, coefficients multiplied to each term. In this paper we use the Jackson kernel
\begin{eqnarray}
	\gamma_n^M = \frac{(M-n+1)\cos\frac{\pi n}{M+1}+\sin \frac{\pi n}{M+1}\cot \frac{\pi}{M+1}}{M+1}.
\end{eqnarray}
It puts most of the weight on the smallest order terms, and the actual number of terms that contribute to the final result is much smaller than $M$.

When calculating the density of states, we are using the second type of expansion since its Chebyshev moments $\mu_n$ are easier to be expressed with MPO, as shown in the main text.

\section{Cutting off of DOS}
%%%%%%%%%%%%%%%

In order to probe thermalization, we are interested in $g( E; \Id)$ and $g( E; \mathcal{O})$ through the full spectrum $[ E_{ \text{min}}, E_{ \text{max}} ]$. But this is challenging; since the DOS is Gaussian, it varies by various orders of magnitude as we vary the energy. To access the tails of the spectrum with the maximum accuracy possible, we proceed piecewise. To do so, we have constructed a step operator $\theta(H-E_{\text{th}})$ that projects $H$ in different energy ranges. The Chebyshev expansion of the first type is particularly convenient for this construction. Consider the step function:
\begin{eqnarray}
\theta(E-E_\text{th})=\left\{
\begin{aligned}
&0,&E<E_\text{th};\\
&1, &E\ge E_\text{th}.
\end{aligned}
\right.
\end{eqnarray}
The corresponding first type Chebyshev moments read
\begin{eqnarray}
\alpha_n^{\theta}=\left\{
\begin{aligned}
&\arccos \tilde{E}_\text{th}/\pi,&n=0;\\
&\sin(n\arccos \tilde{E}_\text{th})/n\pi, &n\ge 1.
\end{aligned}
\right.
\end{eqnarray}
We can use these data to promote the $\theta$ function to a projecting operator, \emph{i.e.},
\bed
\theta(H-E_\text{th}) \approx  \gamma_0^{R}\alpha_0^{\theta} + 2\sum\limits_{m=1}^{R-1}\gamma_m^{R}\alpha_m^{\theta}T_m(\tilde{H}) \\
\eed
\beq
\equiv \theta_R(H-E_\text{th}).
\eeq
Now, given any operator $\opO$, the corresponding truncated DOS with two cutoff parameters $M$ and $R$, $\tr [ \delta_M( \tilde{H} - E ) \theta_R(H-E_\text{th}) \mathcal{O} ]/d_{\mathcal{H}}$, can be estimated from the moments $\mu_n(H; \theta_R(H-E_\text{th}) \mathcal{O})$. When evaluating these moments, products of the form $T^{(D)}_n(H) T^{(D)}_m(H)$ appear.

In principle, it is possible to store all the MPO approximations $T_n^{(D)}(H)$ along the calculation and compute the corresponding cross products, but it is memory consuming. Instead, the computation can be simplified using a simple strategy. Exploiting the relation 
\begin{eqnarray}
T_n(x)T_m(x) = \frac{1}{2}[T_{n+m}(x)+T_{n-m}(x)],
\label{eq:TnTm}
\end{eqnarray}
(where, without loss of generality, we have assumed $n>m$) any product $T_n(H)T_m(H)$ can be expressed as a linear combination of $M+R-1$ Chebyshev polynomials of $H$. This allows us to express the moments of the projected expansion in terms of exactly the same traces as for the unprojected one, but in exchange requires to approximate larger order polynomials (which, as we saw, has an exponential cost in $D$). A more efficient alternative exists to reorganize the computation of the products by invoking again \eqref{eq:TnTm}. Namely, the computation of $T_{n+m}$ can be obtained from the product $T_{\frac{n+m}{2}} T_{\frac{n+m}{2}}$ (for even $n+m$) or $T_{\frac{n+m+1}{2}}T_{\frac{n+m-1}{2}}$ (if $n+m$ is odd).

We can finally express, for even $n+m$,
\begin{eqnarray}
T_n(x)T_m(x) = & T_{\frac{n+m}{2}}(x)T_{\frac{n+m}{2}}(x) \nn\\
+ &\frac{1}{2}[T_{n-m}(x)-T_{0}(x)]
\label{eq:TnTm_even}
\end{eqnarray}
and for odd $n+m$,
\begin{eqnarray}
T_n(x)T_m(x) = & T_{\frac{n+m+1}{2}}(x)T_{\frac{n+m-1}{2}}(x) \nn \\
+ &\frac{1}{2}[T_{n-m}(x) -T_{1}(x)].
\label{eq:TnTm_odd}
\end{eqnarray}
So that we can evaluate the projected expansion if we approximate polynomials up to order $\max(M,R)$, without the need to store them in memory. The only additional step in the algorithm is, at each order $m$, evaluating the corresponding traces for $T_m^{(D)}(H)\ T_m^{(D)}(H)$ and $T_m^{(D)}(H) \ T_{m-1}^{(D)}(H)$, before discarding $T_{m-1}^{(D)}(H)$.

Let us now show how implementing $\theta( H - E_\mathrm{th} )$ helps us estimate $g(E; \mathcal{O})$ on $[E_\text{min}, E_\text{max}]$. We discussed above why the tails of the distribution are problematic. For the sake of concreteness, let us focus on the right end of spectrum, $[E_\text{cut}, E_\text{max}]$, where $E_\text{cut}$ is some threshold value chosen so that $g(E; \Id)$ is monotonically decreasing on $[E_\text{cut}, E_\text{max}]$. We are going to evaluate $g(E; \mathcal{O})$ in a succession of intervals which union is $[E_\text{cut}, E_\text{max}]$. We proceed as follows: 

\begin{itemize}

\item[1] 
Choose some reduction factor $\eta$ in $(0, 1)$.

\item[2] 
Use Chebyshev expansions to provide a non-truncated initial estimate $g_\text{ini}(E;\Id)$ and $g_\text{ini}(E;\mathcal{O})$ respectively for $g(E;\Id)$ and $g(E;\mathcal{O})$ in $[E_\text{min}, E_\text{max}]$. Set $g_0(E;\Id)=g_\text{ini}(E;\Id)$, $g_0(E;\mathcal{O})=g_\text{ini}(E;\mathcal{O})$, $E_0 = E_\text{cut}$ and $s = 0$.

\item[3] 
Find the largest $E_{s+1}$ in $[E_s, E_\text{max}]$ such that $g_s(E_{s+1};\Id) \geq \eta g_s(E_s;\Id)$. $g_s(E;\mathcal{O})$ and $g_s(E;\Id)$ are our estimates respectively for $g(E;\mathcal{O})$ and $g(E;\Id)$ on the interval $[E_s, E_{s+1}]$.

\item[4] Compute $g_{s+1}(E;\Id)=\theta(E-E_{s+1}-\delta)  g_0(E;\Id)$ and $g_{s+1}(E;\mathcal{O})=\theta(E-E_{s+1}-\delta)  g_0(E;\mathcal{O})$, where $\delta$ is a small safety parameter. These functions are constructed from the moments $\mu_n(H;\theta(H-E_{s+1}-\delta))$ and $\mu_n(H;\theta(H-E_{s+1}-\delta)\mathcal{O})$.

\item[5]
$ s \leftarrow s + 1$.

\item[6]
If $E_s < E_\text{max}$, got to 3; else go to 7.

\item[7]
End.
\end{itemize}

The strategy just exposed \emph{does} lead to a more accurate estimate of generalized DOS near the edges of the energy spectrum, as can be appreciated on Fig.~\ref{fig:Ising_therm}.\footnote{We have also checked that the DOS is more accurately estimated in the example of the integrable Ising model.} 
Other strategies may be used to deal with the edges of the energy spectrum, e.g. one could adapt the energy truncation step introduced in~\cite{Holzner2011cheMPS}
to try to suppress the undesired regions.

\begin{figure}
	\centering
	\subfloat[]{\label{fig:err_TnDOS_L80}\includegraphics[width=.5445\textwidth]{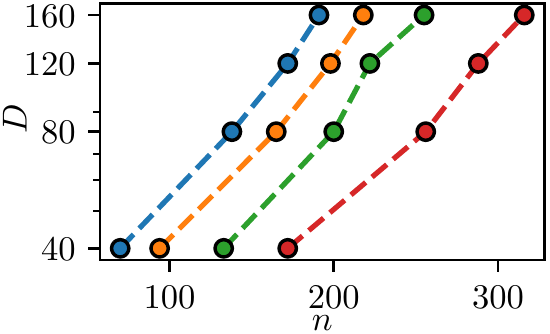}}
	\subfloat[]{\label{fig:err_TnLDOS_L80}\includegraphics[width=.4455\textwidth]{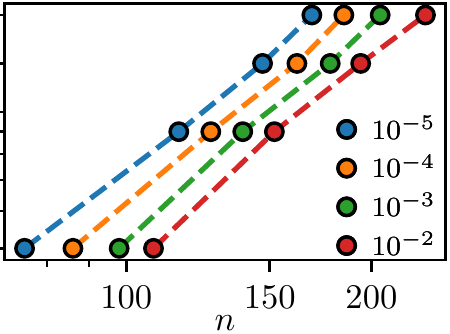}}\\
	\subfloat[]{\label{fig:err_TnDOS_L40}\includegraphics[width=.5445\columnwidth]{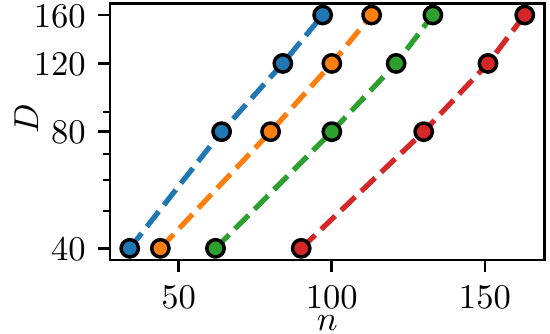}}
	\subfloat[]{\label{fig:err_TnLDOS_L40}\includegraphics[width=.4455\columnwidth]{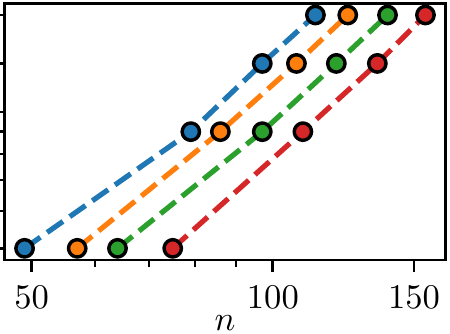}}
	\caption{
		(a)(c): Semi-log plot of bond dimension $D$ required to keep a given truncation error of $10^{-2},10^{-3},10^{-4},10^{-5}$ in the MPO approximation of $T_n(\tilde{H})$ as a function of n for the non-integrable Ising chain, $D_\mathrm{max}=200$; (b)(d): Same but for $T_n(\tilde{H})\Xp$ in a log-log plot. (a)(b): $N=80$; (c)(d): $N=40$. 
	}
	\label{fig:errors}
\end{figure}

\begin{figure}
	\centering
	\includegraphics[width=1.0\columnwidth]{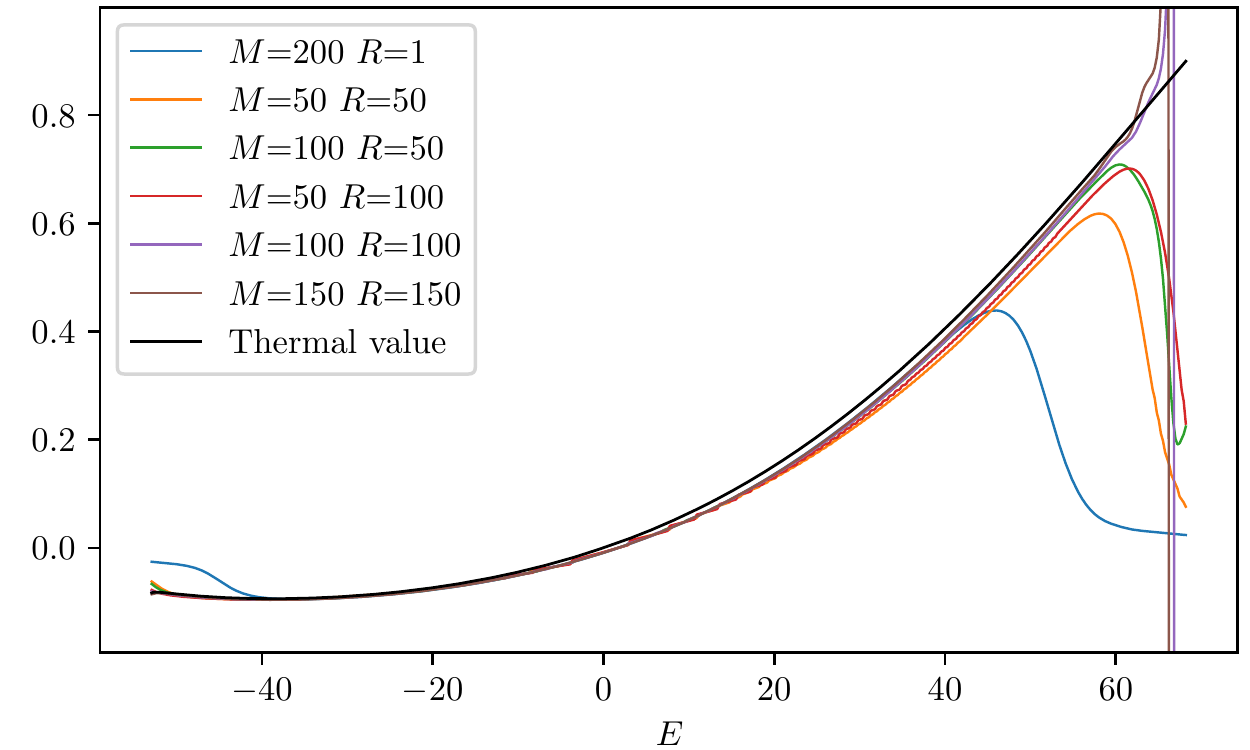}
	\caption{Full spectrum version of upper left plot in Fig.~\ref{fig:Ising_therm}, with more different values of $(M,R)$.}
	\label{fig:thermal_Ising_NI_nocut}
\end{figure}

\section{Error analysis}
%%%%%%%%%%%%

There are two distinct sources of errors in the scheme presented in the main text. The first is that induced by cutting off Chebyshev expansions to some finite order $M$. It is proved~\cite{Weisse2006kpm} that with the Jackson kernel, if $f$ is continuous in $[-1,1]$, the finite sum of the first $M$ terms $f_M$ will converge uniformly to $f$ as
\begin{eqnarray}
||f-f_M||_{\infty} \sim O(1/M).
\label{ChebyErr}
\end{eqnarray}
This error scaling may not be apparent for small spin systems, for which the function to approximate, \emph{i.e.} the spectrum, is discrete. In such cases, increasing $M$ will eventually make this discrete nature emerge. In turn, in the thermodynamic limit, the DOS and related functions become smoother and the results will converge with $M$ as prescribed by (\ref{ChebyErr}).
The second source of error is the bond dimension truncation that takes place after each application of the recurrence relation.
%of the MPO that represent the polynomials, or the MPS that represent their action on the initial state, in the case of the LDOS.
To estimate it numerically, at each step of the iteration we compute the (Hilbert-Schmidt) distance between the best approximated polynomial (corresponding to $D_{\mathrm{max}}$) and its truncations to $D<D_{\mathrm{max}}$,
\beq
\epsilon_n( D ) = \frac{ \| T_n^{(D_{\mathrm{max}})}(\tilde{H}) - T_n^{(D)}(\tilde{H}) \|^2_2}{ \| T_n^{(D_{\mathrm{max}})}(\tilde{H}) \|_2^2 }.
\label{eq:truncT}
\eeq
From these data, we can estimate the bond dimension required to keep the truncation error below a certain threshold. We've made a check on the needed bond dimension $D$ to obtain $T_n(H)$ within a certain error in Fig.~\ref{fig:errors}. As illustrated in Fig.~\ref{fig:errors}, we find this error to grow faster than polynomially with $n$~\footnote{Notice that this procedure gives an estimate of how fast the truncation error grows, but cannot be considered a bound for the error. A stricter bound could in principle be obtained by estimating and adding the error at each truncation step, i.e. the distance between the truncated polynomial $T_{ n + 2 }^{(D)}( \tilde{H} ) $ and the full sum $2 \tilde{H} T^{(D)}_{n+1}(\tilde{H}) - T^{(D)}_{n}(\tilde{H})$.}.

Surprisingly, for the system sizes we have considered, $N=40$ and $N=80$, we have found that to achieve a given error, the bond dimension $D$ required is smaller for the larger system.

In the case of the LDOS calculation, the truncation does not happen at the operator level, but it takes place on the MPS resulting from the application of the polynomials to the initial state. The analogous error,
\beq
\eta_n( D ) =\frac{\| \ket{t_n^{(D_{\mathrm{max}})}}-\ket{t_n^{(D)}}\|^2}{\|\ket{t_n^{(D_{\mathrm{max}})}} \|^2}
\label{eq:truncTPsi}
\eeq
reveals that the effort in $D$ is much more modest for these quantities. Our results suggest (see right panel in fig.~\ref{fig:errors}) a polynomial scaling of $D$ as a function of $n$. This latter observation is in line with the conclusions of \cite{Holzner2011cheMPS}, while estimating other functions, such as the full DOS, appears to be qualitatively more demanding.

When probing thermalization, \emph{i.e.}, when calculating $g(E;\mathcal{O})/g(E;\Id)$, the error results mainly come from the interplay among three types of cutoffs : finite MPO bond dimension ($D$), finite order for the Chebyshev expansion of (weighted) DOS ($M$), and finite order for the Chebyshev expansion of the $\theta$ function ($R$). If we focus on the bulk of the spectrum, even if the error induced by finite $D$ is non-negligible, increasing the sum of $M + R$ will result in an improvement comparable to that obtained if $D$ had been taken large enough that the MPO error can be ignored. But at the edges, the error soon diverges or oscillates quite quickly. Since the DOS is not dense in the tail, we believe we can deal with this lack of accuracy by cutting 2\%-3\% of the edges on both sides in the thermal property plots of Fig.~\ref{fig:Ising_therm} and \ref{fig:PXP_th}.

We have also observed that near the edge of the spectrum, a reconstructed DOS (or related function) may 'spill' beyond the minimum / maximum energy; using finite order Chebyshev expansions inevitably produces some broadening. This effect has a significant impact on the LDOS and the surviving probability; since the weight function $1/(\pi\sqrt{1-x^2})$ is very large near the edges, such out-of-bounds contributions are strongly amplified. To counter this effect, we have rescaled and shifted the spectrum to make it fit into some interval $[-1+\epsilon,1-\epsilon]$~\cite{Holzner2011cheMPS}, where $\epsilon$ is some safety parameter. And in the final step of LDOS calculations (getting $g(E;\Psi)$ from $\mu_n( H; \Psi)$), the range of energy involved has been extended a bit, for instance, to $\left[-1+\epsilon/2,1-\epsilon/2\right]$.
\\

\section{Survival Probability of Ising models}

The lower panels of fig.~\ref{fig:Ising_therm} show $F_M(t)$ as a function of time for the three initial states and both Ising models, for system size $N=40$, up to times where the order truncation becomes significant (recognized by comparing different values of $M$). We observe qualitative differences among the states, with the $\Yp$ state, always supported by interior energy eigenstates, exhibiting a faster decay, whereas the other states, which lie closer to the edges of the spectrum, survive longer. The plot indicates that the required Chebyshev truncation order $M$  grows faster than polynomially with time. Our findings are compatible with the features predicted in e.g. \cite{Tavora2016power,Santos2018}, such as the survival collapse and the onset of the algebraic decay in some cases. 

\begin{figure}[h]
	\centering
	\begin{minipage}{.494\columnwidth}
		\includegraphics[width=\columnwidth]{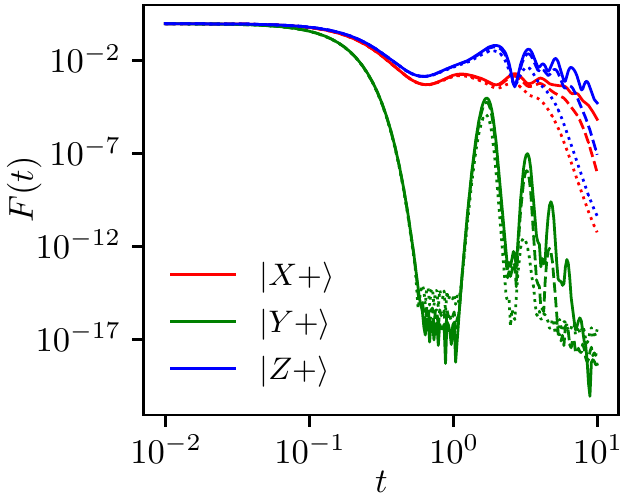}
	\end{minipage}
	\begin{minipage}{.494\columnwidth}
		\includegraphics[width=\columnwidth]{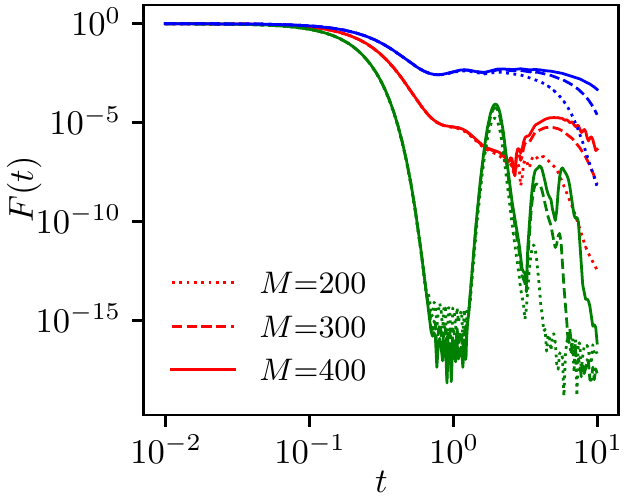}
	\end{minipage}
	\caption{
		Survival probability of $\Xp$, $\Yp$ and $\Zp$ as a function of time of the non-integrable (left) and integrable (right column) Ising models, for a chain of $N=40$ sites. Different line styles corresponds to different truncation order in the LDOS.
	}
	\label{fig:Ising_SP}
\end{figure}

%% file: appendix-degenerate.tex
\section{Degenerate spectrum}

If the spectrum is degenerate, we can still use the Chebyshev expansions to estimate the long-time averaged limit of expectation values.
We can write the evolved state as
\beq
\ket{\Psi(t)}=\int dE e^{-i E t}\sum_k \delta(E-E_k) \ket{k}\Psi\rangle,
\eeq
where the sum over $k$ runs over all energy eigenstates.
The time averaged expectation value of $\opO$ can then be written as
\begin{widetext}
\begin{align}
\bar{O}=&\frac{1}{T}\int dt \bra{\Psi(t)} \opO \ket{\Psi(t)} 
=\frac{1}{T}\int dt \int dE \int dE' \sum_{k,k'}\delta(E-E_k)\delta(E'-E_k')e^{-i (E-E')t} \bra{\Psi} k'\rangle \bra{k'}\opO\ket{k} \bra{k} \Psi\rangle \nn \\
=&\int dE \sum_{k,k'} \delta(E-E_k)\delta(E-E_{k'}) \bra{\Psi} k'\rangle \bra{k'}\opO\ket{k} \bra{k} \Psi\rangle \equiv \int dE \bar{O}(E),
\label{eq:intO}
\end{align}
\end{widetext}
where we have integrated the time already, so that only terms with $E=E'$ survive.
Now if we focus on the argument of the energy integral, we can expand each of the delta functions as we have done before in terms of Chebyshev polynomials,
\beq
\delta(E-E_k)\approx \frac{1}{\pi \sqrt{1-E^2}}\sum_m c_m T_m(E_k) T_m(E),
\eeq
(where for simplicity, we are assuming that the Hamiltonian is already rescaled). Inserting this expansion (twice) in \eqref{eq:intO}, we find

\begin{widetext}
\begin{align}
\bar{O}(E)=& \frac{1}{\pi^2 (1-E^2)}\sum_{m,p}c_m c_p T_m(E) T_p(E) \sum_{k,k'}
 \bra{\Psi} k'\rangle \bra{k'} T_p(E_{k'}) \opO T_m(E_k) \ket{k} \bra{k} \Psi\rangle\nn\\
 =& \frac{1}{\pi^2 (1-E^2)}\sum_{m,p}c_m c_p T_m(E) T_p(E)  \bra{\Psi}T_p(H) \opO T_m(H) \ket{\Psi}.
\end{align}
\end{widetext}

To evaluate the above expression, we simply compute the vectors $\ket{t_m}$ as in the main text, and use them to evaluate the corresponding matrix elements of $\opO$.
Notice that if there was no degeneracy, $k$ and $k'$ will be the same, and we recover the diagonal ensemble form presented in the main text.